\documentclass[oldversion,letter]{aa}
\usepackage{txfonts}
\usepackage{graphicx}
\usepackage[]{natbib}
\usepackage{mathrsfs}
\bibpunct{(}{)}{;}{a}{}{,}

\newcommand{\chan}[0]{\textit{Chandra}}
\newcommand{\xmm}[0]{XMM-\textit{Newton}}
\newcommand{\bd}[0]{\mbox{2MASS~0535-0546}}

\begin{document}

\title{Discovery of X-ray emission from the eclipsing brown-dwarf binary \mbox{2MASS J05352184-0546085}}
\author{S. Czesla \and P. C. Schneider \and J. H. M. M. Schmitt}
\institute{Hamburger Sternwarte, Universit\"at Hamburg, Gojenbergsweg 112, 21029 Hamburg, Germany}
\date{Received ... / accepted ...}
\abstract
{
  The eclipsing brown-dwarf binary system \mbox{2MASS J05352184-0546085} is a case sui generis. For the first time, it 
  allows a detailed analysis of the individual properties of young brown dwarfs, in particular, masses, and radii, and the temperature
  ratio of the system components can be 
  determined accurately. The system shows a ``temperature reversal" with the more massive component being
  the cooler one, and both components are found to be active.
  We analyze X-ray images obtained by \chan~ and
  \xmm~ containing \mbox{2MASS J05352184-0546085} in their respective field of view.
  The \chan~ observatory data show a clear X-ray source at the position of \mbox{2MASS J05352184-0546085}, whereas the \xmm~ data suffer from contamination from other nearby sources, but are consistent with the \chan~ detection.
  No indications of flaring activity are found
  in either of the observations (together $\approx 70$~ks), and we thus attribute the observed flux to quiescent emission.
  With an X-ray luminosity of
  $3\times10^{28}$~erg/s we find an \mbox{$L_X/L_{bol}$}-ratio close to the saturation limit of $10^{-3}$ and
  an \mbox{$L_{X}/L_{H\alpha}$}-ratio consistent with values obtained from low-mass stars.
  The X-ray detection of \mbox{2MASS J05352184-0546085} reported here provides additional support for the interpretation of the temperature 
  reversal in terms of magnetically suppressed convection, and suggests that the activity phenomena of young brown dwarfs
  resemble those of their more massive counterparts.
}
\maketitle

 \section{Introduction}
   Magnetic activity is extremely common among late-type stars with outer convection
   zones \citep[e.g.,][]{schmitt_1997}. The cause of that activity as diagnosed by chromospheric and coronal emissions is ultimately thought to be a magnetic dynamo operating near the
   base of the convection zones of these stars. Near the spectral type $\sim$~M3, stars are thought to become fully convective, thereby
   preventing the emergence of a solar like dynamo in their interiors, and yet
   the activity properties seem to remain unchanged across this borderline \citep[e.g.][]{fleming_1993}.
   A saturation level of
   \mbox{$L_X/L_{bol} \sim 10^{-3}$} is observed also for fully convective stars, although with substantial dispersion.
   Ultra cool dwarfs (spectral type M7 and later) show a drop in activity
   as measured through their H$_{\alpha}$ emission, and among L-type dwarfs
   H$_{\alpha}$ emission is rarely found \citep{gizis_2000}.
   For the majority of active M dwarfs the ratio between H$_{\alpha}$ and bolometric
   luminosity is \mbox{$\log(L_{H\alpha}/L_{bol}) \sim -3.8$},
   while for most objects of spectral class~L this ratio drops below $-5$. \citet{mohatny_2002} argue that the high degree of
   neutrality in the outer atmospheres of very low-mass stars effectively diminishes the activity level with effective temperature.

   X-ray detections from ultra cool dwarfs are quite sparse, 
   and only a few objects have been identified as X-ray sources during flaring and 
   quiescent periods. Examples comprise the
   binary system \mbox{GJ 569 B} consisting of two evolved (age $\approx 500$~Myrs) substellar objects of late-M spectral type
   detected by \citet{stelzer_2004} and the latest (with respect to spectral type) object being the M9 dwarf 
   LHS~2065 \citep[e.g.,][]{schmitt_2002, robrade_2008}.
   Hence, the behavior of 
   X-ray emission is unclear as one approaches the boundary towards the substellar 
   brown dwarfs and also the role played by the different stellar parameters for the 
   activity evolution of these objects remains unclear \citep{stelzer_2007}.
   It is particularly surprising that some of the very rapidly rotating substellar 
   objects show little or no signs of activity.
   During the very early phases of their evolution, however, when the optical
   characteristics of brown dwarfs
   resemble those of stars with spectral type M6 or so, they are often found to be quite 
   active. Starting with the first X-ray detection of \mbox{Cha H$\alpha$ 1} in the 
   Chamaeleon~I star forming region \citep{neuhauser_1998}, quite a number of
   brown dwarfs settled in star forming regions and (young) clusters have been detected in X-rays.
   The detection rate is higher for young brown dwarf compared to
   evolved ones, which might be related to a higher intrinsic X-ray
   luminosity, hotter outer atmospheres, or both.

   A crucial advance in the observational endeavor to unravel the nature of low-mass objects was the discovery
   of the eclipsing brown-dwarf
   binary \citep[``2MASS J05352184-0546085" - in the following \bd,][]{stassun_2006}, which allowed to directly measure masses, radii,
   and a temperature ratio for brown dwarfs for the first time.  
   The \bd~system is thought to be a member of the
   Orion Nebular Cluster (ONC) and, hence, very young (age $\approx 10^{6}$~years). 
   It consists of two \mbox{$M6.5\pm0.5$} dwarfs in an eccentric ($e = 0.31$) $9.78$~day orbit, such that 
   the two eclipses are separated by a phase difference of $0.67$.  In their analysis of \bd~
   \citet{stassun_2006} determine
   masses of \mbox{M$_1=56\pm4$~M$_{Jup}$} and \mbox{M$_2=36\pm3$~M$_{Jup}$} as well as radii of \mbox{R$_1=0.67\pm0.03$~R$_{\sun}$}
   and \mbox{R$_2=0.49\pm0.02$~R$_{\sun}$} for the primary and secondary component, respectively.
   Astonishingly, a ``temperature reversal" is observed
   in this system, i.e., the higher-mass primary component was found to be cooler than its lower-mass
   companion \citep[T$_1\approx 2700$~K and T$_2 \approx 2800$~K,][]{stassun_2007, reiners_2007}.

   From the activity point of view \bd~is particularly interesting since both components appear to be substantially
   active, and according to \citet{reiners_2007} the H$_{\alpha}$-emission of the primary exceeds that of 
   the secondary by a factor of 7.
   The same authors find a rotational velocity of about $v\sin(i)=10$~km/s for the primary, while 
   they give only an upper limit of 5~km/s for the secondary. Assuming a magnetic origin of this activity 
   and the same scaling between \mbox{H$_{\alpha}$-emission} and magnetic field strength as found for earlier type 
   stars, \citet{reiners_2007} argue
   for the presence of magnetic fields of $Bf\approx 4$~kG for the primary and $Bf\approx 2$~kG for the secondary.

   In this letter we report the discovery of X-ray emission of the eclipsing brown-dwarf binary \bd, 
   which provides further support for the presence of magnetic activity in this system.

  \section{Observations and data analysis}
   \bd~was serendipitously covered by observations with both the \chan~ and \xmm~ observatories; the
   basic properties of these observations are provided in Table~\ref{tab:Obs}.

    \begin{table}[h!]
    \begin{minipage}[h]{0.5\textwidth}
    \renewcommand{\footnoterule}{}
      \caption{X-ray data of \bd.
      \label{tab:Obs}}
      \begin{tabular}{l l l l} \hline \hline
        Instr. & ID   & Obs. time      & Obs. date \\ 
               &      &  [ks]          & [MJD]           \\ \hline
        \chan~ & 2548 & 48.5           & 52523.54        \\
        \xmm~ & 0112660101 & 22.1      & 52167.08 \\ \hline
      \end{tabular}
      \vspace{-3mm}
    \end{minipage}
    \end{table}

  \subsection{The \chan~data \label{sect:chand_data}}
\begin{figure}[h]
      \centering
       \setlength\fboxsep{0pt}
       \setlength\fboxrule{0.5pt}
              \includegraphics[width=0.45\textwidth]{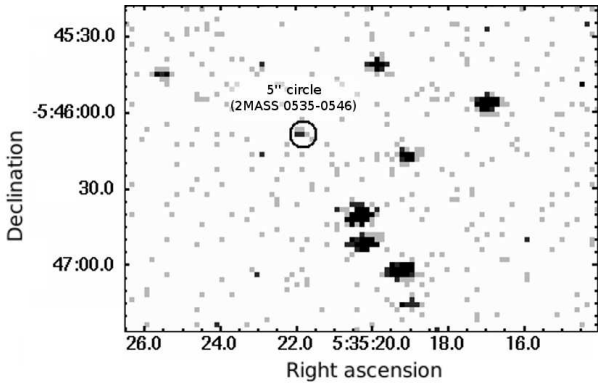}
      \caption{X-ray image ($2$~arcsec/bin) of the surrounding of \bd. A circular (5~arcsec radius) region centered on the
      nominal target position is superimposed on the image. 
      \label{fig:SpecSur}}
    \end{figure}

    The ``flanking field south" of the ONC containing \bd~ was observed with the ``ACIS~I" instrument onboard
    \chan. In the associated image \bd~is located relatively far off-axis \mbox{($6.7$~arcmin)}, but is still well
    exposed, so that a sensitive search for X-ray emission is feasible.
    The \chan~image in the vicinity of \bd~is shown in Fig.~\ref{fig:SpecSur}.
    
    In an effort to suppress background and with the knowledge, that young brown dwarfs have X-ray spectra with
    median energies of the order of or higher than $1$~keV \citep[e.g.][]{preibisch_2005},
    we limited the energy range of the image to the $0.5-2.5$~keV band;
    a circle of radius 5~arcsec is drawn around the nominal position of \bd. 
    According to our point-spread-function (PSF) modeling the encircled region 
    contains $\approx 95$\% of the $1$~keV photons
    from a point source at the respective off-axis position of the target.
    Within this circle we find a total of
    8 photons. Extrapolating the photon numbers measured in the same energy band from nearby source-free regions yields
    an expectation value of 0.8 background counts. The Poisson probability of obtaining 8 or more counts with an expectation value
    of 0.8 is only $2\times 10^{-7}$. Therefore, we attribute the recorded signal to
    an X-ray source with a count level of $7.2\pm2.8$~photons.
    What about out-of-band photons? In the considered source region not a single photon with an energy below 
    $500$~eV was detected,  while two photons
    are contained in the $2.5-10$~keV band. This number is, however,
    in good agreement with the expectation value derived from the close-by
    comparison region, and we, thus, attribute these photons to background.
    
    The median energy of the source photons is $1.4$~keV, a value going well with those obtained
    by \citet{preibisch_2005} for 8~other brown dwarf members of the ONC detected in a quiescent state.
    For the depth of the absorbing column towards \bd~we estimated a value of \mbox{$n_H = 10^{21}$~cm$^{-2}$}
    combining the visual extinction of $A_V\approx 0.75$~mag given by \citet{stassun_2006} with 
    the relation
    $$ n_H\left[\mbox{cm}^{-2}\right] = 1.79\times 10^{21}~\mbox{mag}^{-1} \cdot A_V $$
    derived by \citet{predehl_1995}; a result consistent with
    the lower values derived from brighter, close-by X-ray sources.
    Assuming an absorbed one-component thermal spectrum with subsolar ($0.3$) abundances,
    a plasma temperature of $\approx 2$~keV,
    and a distance of $450$~pc \citep[e.g.][]{stassun_2006} we compute an X-ray luminosity
    of \mbox{$(3.0\pm 1.2)\times 10^{28}$~erg/s} in the $0.5-2.5$~keV band, which -- again -- is
    well covered by the range determined by \citet{preibisch_2005}.
    
    To check the uniqueness of our identification we searched for other potential emitters close (30~arcsec) 
    to \bd, but neither Simbad\footnote{http://simbad.u-strasbg.fr/simbad/}, NED\footnote{http://nedwww.ipac.caltech.edu/}, nor 
    the 2MASS-survey provided any candidates in the environment under consideration.    Therefore, we attribute the detected 
    X-ray emission to \bd.
    
   \subsubsection{Timing}

\begin{figure}[h]
      \centering
      \includegraphics[height=0.45\textwidth, angle=-90]{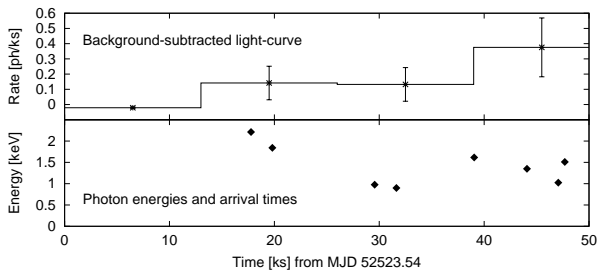}
      \caption{Upper panel: Background-subtracted $0.5-2.5$~keV band light curve ($13$~ks binning). Lower
      panel: Photon energy against arrival time.
      \label{fig:Timing}}
    \end{figure}

    Do the source photons arrive homogeneously distributed in time or not? In Fig.~\ref{fig:Timing}
    we present the background-subtracted $0.5-2.5$~keV band light-curve of \bd~in the upper panel
    and the individual photons (registered by their energies) plotted against arrival time in the lower panel.
    While no photons arrived during the first quarter of the observation, the assumption of a constant arrival rate
    is still consistent with the observations.  More importantly, the light curve (cf., Fig.~\ref{fig:Timing}) suggests
    the presence of persistent X-ray emission more or less homogeneously distributed in arrival time and not
    flaring emission as -- for example -- observed in a number of brown dwarf ONC members \citep{preibisch_2005} or the ultra cool 
    dwarf LHS~2065 \citep{schmitt_2002}.

  \subsection{The \xmm~data}
    As \bd~ is clearly detected in the \chan~ data, we searched for further evidence in other serendipitous
    X-ray observations. In an \xmm~ observation of $\iota$~Orionis \bd~ is 
    located roughly 9~arcmin off-axis and we perform a similar analysis as for the \chan~ data, i.e.,
     we define a source-free background region in the vicinity of \bd~ and concentrate on the energy range $0.5-2.5$~keV. We calculate the excess counts in a circular region with a radius of 15~arcsec (to take into account the larger \xmm~ PSF) around the nominal source position.

    While in the \chan~ image \bd~ is clearly separated from all neighboring sources 
    (see Fig. \ref{fig:SpecSur}), we can unfortunately not exclude some contamination 
    from close-by sources in the \xmm~ data.  Furthermore, in the image constructed 
    from the pn data (i.e., the most sensitive instrument onboard \xmm) the readout 
    strip generated by the out of time events from the central bright source 
    ($\iota$~Orionis) is located less than one arcmin from \bd~ providing yet an 
    additional source of possible contamination.
    Momentarily neglecting these problems, we find $8 \pm 4$, $7 \pm 4$ and $13 \pm 6$ excess photons above the background for the MOS~1, MOS~2, and pn, data respectively.  Converting into count rates and fluxes using the same model as for the \chan~photons
    (cf. Sect.~\ref{sect:chand_data}) and a vignetting factor of 0.6 (see \xmm-Users Handbook sect.~3.2.2.2), 
    we find a value of \mbox{$(14.0 \pm 4.0) \times10^{28}$~erg/s} for the averaged X-ray luminosity in the MOS~1, MOS~2,
    and pn data using WebPIMMS\footnote{http://heasarc.nasa.gov/Tools/w3pimms.html}. This is a factor of five 
    higher than the X-ray flux from \bd~ observed with \chan. If we now estimate the contamination level of the 
    source region by placing a similarly sized region at a comparable position with respect to the neighboring,
    contaminating sources, the observed flux level decreases to values roughly consistent 
    with the \chan~ measured flux. Although the \xmm~data alone are insufficient to claim a detection, we conclude
    that they are at least consistent with the
    assumption of X-ray emission from \bd~ at the same level as observed with \chan, and from an inspection 
    of the arrival times of individual photons we can also exclude
    the presence of (stronger) flares.

  \subsection{Phase coverage of X-ray observations}
  
\begin{figure}[h]
      \centering
      \includegraphics[width=0.45\textwidth]{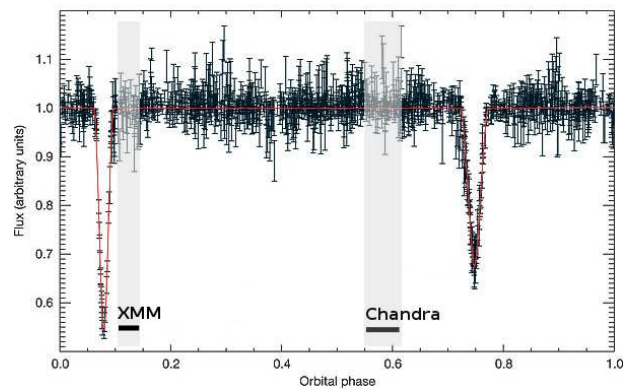}
      \caption{Phase folded photometric data from \citet{stassun_2006}. Observation phases of \chan~and
      \xmm~are indicated by black bars at the bottom and a gray column.
      \label{fig:Phase}}
    \end{figure}

  In Fig.~\ref{fig:Phase} we show the phase folded optical light curve of \bd~ as given by \citet{stassun_2006}, where we  
  indicate those phase intervals covered either by \xmm~or \chan. The \xmm~observation
    takes place shortly after the eclipse of the secondary, but unfortunately, none of the eclipses is covered
    by the presently available X-ray data.  Therefore, we are unable to unambiguously attribute the detected X-ray emission to any
of the individual components of \bd , which is clearly a task for follow-up observations.
    
  \subsection{Chromospheric and coronal activity}
    The bolometric luminosity, $L_{bol}$, of the individual components can be calculated from 
    \begin{equation}
    L_{bol} = 4 \pi r^2 \sigma T^4 \, ,
    \end{equation}
    with $r$ denoting the radius of the brown dwarf, $\sigma$ the Stefan-Boltzmann constant, and $T$ the effective temperature.
    Using the values given by
    \citet{stassun_2007} for the radii and temperatures of $2700$~K and $2800$~K, we find bolometric luminosities of 
    \mbox{$8\times10^{31}$~erg/s} and \mbox{$5.1\times10^{31}$~erg/s} for the primary and secondary components.

    As the primary shows strong magnetic activity \citep{reiners_2007}, we naturally blame it to be also 
    responsible for the X-ray emission.
    Comparing the X-ray luminosity of \mbox{$3\times10^{28}$~erg/s} to its bolometric luminosity leads
    to a ratio of \mbox{$\log L_X/L_{bol} = -3.4$}, and similar results of $-3.2$ and $-3.6$ are obtained 
    if not the primary but the secondary or both were the X-ray bright component(s).
    
    \citet{preibisch_2005} detected quiescent emission from 8~young brown dwarfs in the ONC. We find that
    both the X-ray luminosity of \bd~and the \mbox{$\log L_X/L_{bol}$-ratio} (no matter whether the primary or
    the secondary is held responsible for the X-ray emission) point to a rather active, however, not unprecedented object.
    From the activity point of view, it appears 
    similar to the late M~dwarf \mbox{2MASS J05350705-0525005} (COUP 280), which is the most active brown dwarf in the (quiescent) \citet{preibisch_2005} sample showing an \mbox{$\log L_X/L_{bol}$-ratio} of $-3.3$; with an
    X-ray luminosity of $5\times10^{28}$~erg/s it is also among the most luminous.
    This provides further support for our attribution of the recorded X-ray 
    flux to the \bd~ system.
    
    What do these outcomes imply for the relation between chromospheric and coronal activity in \bd?
    \citet{reiners_2007} derive values of \mbox{$\log(L_{H\alpha}/L_{bol})=-3.5$} for the primary 
    and \mbox{$\log(L_{H\alpha}/L_{bol})=-4.3$} for the secondary. Substituting our results we obtain
    \mbox{$\log(L_{X}/L_{H\alpha})=0.1$} if the primary were the X-ray bright
    constituent and \mbox{$\log(L_{X}/L_{H\alpha})=0.7$} if it is the secondary. 

    These values compare well to those given by \citet{dahm_2007} for similarly young low-mass ($M<0.5\,M_{\sun}$) classical 
    and weak-line T-Tauri stars in NGC~2264,
    and are also within the range of
    \mbox{$\log(L_{X}/L_{H\alpha})$}-ratios obtained by \citet{hawley_1996} for active cluster and field M-dwarfs in a more
    advanced evolutionary state.
 
   \section{Summary and conclusions}
 
   In summary, we find an unambiguous X-ray detection of the eclipsing brown-dwarf binary \bd. 
   An X-ray source at the nominal position of \bd~ is clearly seen with the \chan~observatory, and
   the results are also compatible with an \xmm~image observed about one year earlier.
   The X-ray luminosity of \bd~ is of the order of $3\times 10^{28}$~erg/s, and even though the 
   \chan~ light curve might indicate some moderate increase in luminosity, no flaring activity 
   is seen in either of the observations. Therefore, we attribute the flux to quiescent X-ray emission.
    
   An inspection of the source photons recorded by \chan~shows that the observed plasma emits at a temperature of
   at least $1$~keV and probably more. The relative hardness of the spectrum is compatible with the results 
   obtained by \citet{stelzer_2007} for a wide and similarly young brown-dwarf binary system.
   From the bolometric luminosity of the primary component we compute a value of $-3.4$ for the
   \mbox{$\log(L_X/L_{bol})$}-ratio, 
   which is close to the saturation limit of $\approx -3$ and within the range of values determined by
   \citet{preibisch_2005} for other brown dwarf members of the ONC detected in quiescence as well as compatible with
   values computed by \citet{stelzer_2006} for other young brown dwarfs.
   Invoking the findings of \citet{reiners_2007} we 
   find \mbox{$\log(L_{X}/L_{H\alpha})$}
   values well consistent with previously published results for low-mass stars.
   Unfortunately, none of the optical  eclipses is covered by the available X-ray data, so we have no 
   way at the moment to assign the X-ray flux to the individual components of \bd .
   
   What can our results tell us about the origin of the temperature reversal?
   The effect may be caused either by a non-coeval formation of the brown dwarfs
   or by the impact of magnetic fields, which hamper energy transport by convection.
   The X-ray detection of \bd~ alone supports the presence of magnetic fields causing the 
   X-ray emission; further evidence is the hardness of the source photons indicating that magnetic 
   processes are the source of the activity. Our results are, therefore, fully in line with those 
   of \citet{reiners_2007}, pointing towards the presence of strong magnetic
   activity in \bd~ near the saturation limit and hence a magnetic origin of the observed 
   temperature reversal.
    
  \acknowledgement{This work has made use of data obtained from the \chan~ and \xmm~ data archives.
  S.C. acknowledges support from the DLR under grant 50OR0105.
      P.C.S. acknowledges support from the DLR under grant 50OR703.
  }
  \bibliographystyle{aa}
  \bibliography{0818bibl}
  
\end{document}